\documentclass{iaus}
\usepackage{graphicx}

\title[Feedback and Dark Matter] 
{Effects of SN Feedback on the Dark Matter Distribution}

\author[S.E. Pedrosa, P.B. Tissera \& C. Scannapieco]   
{Susana E. Pedrosa$^1$$^2$, Patricia B. Tissera$^1$$^2$
 \and Cecilia Scannapieco$^3$}

\affiliation{$^1$Consejo Nacional de Investigaciones Cient\'{\i}ficas y T\'ecnicas, CONICET, Argentina \\ email: {\tt supe@iafe.uba.ar} \\[\affilskip]
$^2$Instituto de Astronom\'{\i}a y F\'{\i}sica del Espacio, Casilla de Correos 67, Suc. 28, 1428, Buenos Aires, Argentina \\ email: {\tt patricia@iafe.uba.ar} \\[\affilskip]
$^3$Max-Planck Institute for Astrophysics, Karl-Schwarzchild Str. 1, D85748, Garching, Germany \\email: {\tt cecilia@mpa-garching.mpg.de}}

\pubyear{2008}
\volume{IAU Symposium No.254}  
\pagerange{}
\jname{The Galaxy Disk in Cosmological Context}
\editors{A.C. Editor, B.D. Editor \& C.E. Editor, eds.}
\begin{document}

\maketitle

\begin{abstract}

We use cosmological simulations to study the effects of supernova (SN) feedback on the dark matter distribution in galaxies. We simulate the formation of a Milky-Way type galaxy using a version of the SPH code GADGET2 which includes chemical enrichment and energy feedback by SN, a multiphase model for the gas component and metal-dependent cooling.
We analyse the impact of the main three input SN feedback parameters on the amplitude and shape of the dark matter density profiles, focusing on the inner regions of the halo.
 In order to test the dependence of the results on the halo mass, we  simulated a scale-down version of this system.
First results of this ongoing work show that the dark matter distribution is affected by the feedback, through the redistribution of the baryons. Our findings suggest that the response of the dark matter halo  could be  the result of a combination of several physical parameters such as the amount of stellar mass formed at the centre, its shape, and probably the bursty characteristics of the star formation rate. As expected, we find that the dark matter haloes of small galaxies are more sensitive to  SN feedback. Higher resolution simulations are being performed to test for numerical effects.

\end{abstract}

\firstsection 
\section{Introduction}

Numerical simulations are an important tool for studying galaxy formation and evolution as they are able to describe the non-linear evolution of dark matter and baryons in a consistent way. The study of halo density profiles is particularly interesting because of the possibility of using them  to confront $\Lambda$CDM models with observational data such as disk galaxy rotation curves and the shapes of the cores of dwarf galaxies.
The universality of the dark matter density profiles has been claimed in several works (\cite[Navarro et al. 1995]{Nav95}; \cite[Navarro et al. 1996]{Nav96}; \cite[Navarro et al. 1997]{Nav97}).
The halo profiles have been found to deviate slightly but systematically from the NFW model proposed by \cite[Navarro et al. (1995)]{Nav95} as reported by recent works (\cite[Navarro et al. 2004]{Nav04}; \cite[Prada et al. 2006]{Prada06}; \cite[Merrit et al. 2006]{Merrit06}). \cite[Gao et al. (2008)]{Gao08} and \cite[Merrit et al. (2006)]{Merrit06} showed that the Einasto profile (\cite[Einasto  1965]{Einas65}) provides a better fit to the dark matter halo distribution.

It is believed that the presence of baryons in the central regions
of the haloes can  modify the dark matter potential.
\cite[Blumenthal et al. (1986)]{Blumen86} proposed the adiabatic
contraction model to globally predict the effects of baryons on the
dark matter distribution. However,  this model might be an
oversimplification of the problem and more realistic treatments are
needed. \cite[Choi et al. (2006)]{Choi06} studied the accuracy of
the circular-orbit adiabatic approximation  in predicting halo
contraction due to disk formation. Although they found that the
adiabatic approximation is valid, they did not include  any other
physical mechanism that could modify the halo structures. Baryons
can also affect the distribution of matter through the injection of
SN energy, principally in low mass systems. \cite[Gnedin \& Zhao
(2002)]{Gnedin02} studied the effect of maximum feedback on the
central density of dark matter halos of gas-rich dwarfs, combining
analytical models with simulations, finding that the effect is too
weak to agree with observed rotation curves. \cite[Read \& Gilmore
(2005)]{Read05} found that two impulsive mass-loss phases are needed
to significantly change the dark matter profiles.

Our aim is to  investigate if SN feedback can indirectly affect the
dark matter halo profiles by the effects produced on the baryonic
distributions. In this work, we present preliminary results on the
analysis of hydrodynamical numerical simulations of galaxies formed
in a hierarchical clustering scenario. It is organized as follows.
In Section 2 we present our numerical experiments. In Section 3 we
discuss the preliminary results obtained for the dark matter density
profiles and for the rotation curves in connection with the results
for the dark matter profiles. In section 4 we summarize our
preliminary findings.

\section{Numerical Experiments}

We use cosmological simulations to study the effects of SN feedback on the dark matter distribution in galaxies. We  used an extension of the SPH code GADGET2  developed by \cite[Scannapieco et al. (2005)]{Scab05} and  \cite[Scannapieco et al. (2006)]{Scan06} which   includes chemical enrichment and energy feedback by SN, a multiphase model for the gas component and metal-dependent cooling.
A set of 8 Milky Way type simulations with the same initial condition but different input feedback parameters were run. We adopted a $\Lambda$CDM Universe with the following cosmological parameters: $\Omega_{\Lambda}=0.7$, $\Omega_{m}=0.3$, $\Omega_{b}=0.04$, a normalization of the power spectrum of $\sigma_{8}=0.9$ and $H_{0}= 100 h \ {\rm km} \ {\rm s}^{-1}\ {\rm Mpc}^{-1}$, with $h=0.7$. The particle mass is $1.6\times 10^{7} h^{-1}\ M_{\odot}$ and $2.4\times 10^{6} h^{-1}\  M_{\odot}$ for the dark matter and baryonic particles, respectively.  We have also run a simulation without feedback for comparison.  The gravitational softening used was 0.8 $h^{-1}$ kpc. The halo was extracted from a cosmological simulation and resimulated with higher resolution, and was selected to have no major mergers since z=1. At z=0 the MW halos  are  relaxed as indicated by their relax parameter of $\approx 0.002$ (\cite[Neto et al. 2007]{Neto07}).
Details on these simulations can be found in \cite[Scannapieco et al. (2008)]{Scan08} (hereafter, S08),
 where an extensive study on the effects of SN feedback on the star formation rates,
galaxy morphology and disk formation is presented.

The star formation (SF) and feedback model has three relevant inputs parameters: the feedback parameter, $F$ (which regulates the fraction of energy injected into the cold and hot phases and ultimately determines the feedback strength), the energy released per SN, $E$, and the star formation efficiency, $C$.
In Table 1 we show the main characteristics of the experiments and the input parameters assumed in each case. Depending on the combination of SF and SN parameters, baryons settle down determining structures with different morphologies.
 Disk components of different sizes and masses can be produced. However,
in most cases the systems are dominated  by a spheroidal component ($D/S < 1$).
As discussed in S08, the inclusion of SN feedback allows the formation of disk components as a result of the self-regulation of the star formation activity and the generation of galactic winds. On the contrary, if SN feedback is not considered no disk develops and the final galaxy has a spheroidal shape.  As it can be seen in Table~\ref{tab1}, simulations E-0.7, F-0.5 and F-0.9 have the largest disks, while  E-3 and C-05 runs, as well as the no feedback (NF) run were not able to form disks.

\begin{table}
  \begin{center}
  \caption{Main characteristics of  simulations. The feedback parameters used in each simulation: the feedback parameter ($F$), the energy ($E$) per SN (in units of 10$^{51}$ ergs), and the star formation efficiency ($C$). We also show the total stellar mass of the central galaxy (in units of $10^{10} h^{-1}\ M_{\odot}$),the  total to stellar mass ratio, the disk to spheroid mass ratio, the disk scalelength (in $h^{-1}$ kpc), the $n$ and $r_2$ einasto's parameters and the maximum total circular velocity to virial velocity ratio. $M_t$ and $M_s $ are evaluated at the radius that enclosed $83\%$ of the baryons in the central regions. Bootstraps errors for $n$ (in units of $10^{-3}$) and $r_2$ (in units of $10^{-2}$) are shown within parenthesis. }
  \label{tab1}
 {\scriptsize
  \begin{tabular}{|l|c|c|c|c|c|c|c|c|c|}\hline
{\bf Run} & {\bf Feedback} & {\bf M$_{s}$} &  {\bf $M_t/M_s $} & {\bf $D/S$} & {\bf r$_{d}$} & {\bf $n$} & {\bf $r_{2}$} & {\bf $Vc^{max}/V_{200}$} \\
   &  {\bf param} & &  & & & & & \\ \hline
NF & No feed &  15.6 &  4.0 & - & - & 6.262 (4)& 17.39 (1) & 1.39 \\ \hline
F-0.3 & F0.3, E1, C0.1  & 4.5 &  9.7 & 0.42 & 5.8 & 7.037 (6)& 19.13 (1)& 1.18 \\ \hline
F-0.5 & F0.5, E1, C0.1  & 5.5 & 7.4 & 0.82 & 6.5 & 7.659 (5)& 17.86 (2) & 1.20 \\ \hline
F-0.9 & F0.9, E1, C0.1  & 6.6 &  7.7 & 1.04 & 9.7 & 7.101 (1)& 18.34 (1)& 1.22 \\ \hline
E-0.7 & F0.5, E0.7, C0.1 & 7.5 &  5.1 & 0.82 & 5.7 & 7.807 (1)& 15.45 (0.1)& 1.30 \\ \hline
E-0.3 & F0.5, E0.3, C0.1  & 13.5  & 3.8 & 0.60 & 4.8 & 7.053 (3)& 15.31 (1) & 1.44 \\ \hline
E-3 & F0.5, E3, C0.1  & 1.3 & 37.1 & - & - & 6.216 (1)& 23.09 (2)& 1.13 \\ \hline
C-0.01 & F0.5, E1, C0.01 & 9.9  & 3.7 & 0.39 & 2.6 & 7.163 (2)& 16.42 (1)& 1.40 \\ \hline
C-0.5 & F0.5, E1, C0.5  & 4.5  & 11.5 & - & - & 7.998 (1)& 17.15 (0.1)& 1.19 \\ \hline
  \end{tabular}
  }
 \end{center}
\vspace{1mm}
\end{table}

\section{Dark matter density profiles}

We constructed the dark matter profiles for the simulated systems at
$z=0$ and tried to fit them several profile fitting formulae (NFW,
Jaffe, Einasto). All analysed dark matter haloes have more than
120000 particles within the virial radius. Note that we have not
taken out the substructure within the virial radius. Although we are
currently addressing this point we do not expect a significant
change since all our haloes are relaxed as already mentioned. The
best fit was given by the Einasto's expression (\cite[Einasto
1965]{Einas65}) in agreement with previous works. The Einasto
profile has  two free parameters, $n$ and $r_2$, which indicate the
sharpness and maximum position of the curves, respectively. Larger
$n$ values represent more concentrated profiles (\cite[Gao et al.
2008]{Gao08}).
 The fitting values obtained for the different simulations are shown in Table \ref{tab1}. We estimated bootstrap errors for $n$ and $r_2$ by fitting the Einasto's formula to 50 bootstrap dark matter profiles and estimating the standard dispersion over the generated set of $n$ and $r_2$ parameters.
As we can see the shape of the dark matter profiles can change significantly in comparison to the NF case, depending on the
effects of the SN feedback.

\begin{figure}[h]
\begin{center}
 \includegraphics[width=2.4in]{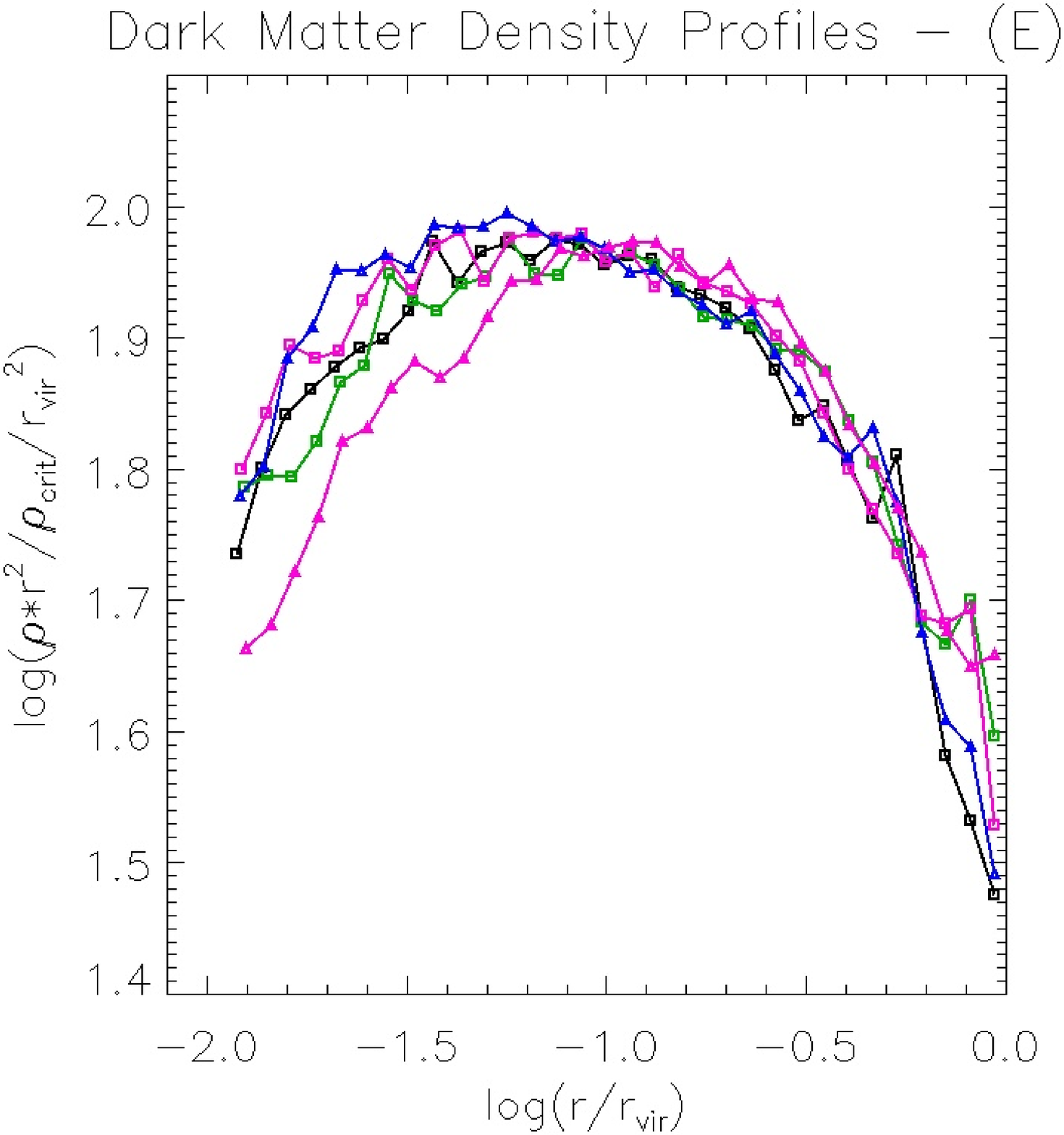}
 \includegraphics[width=2.4in]{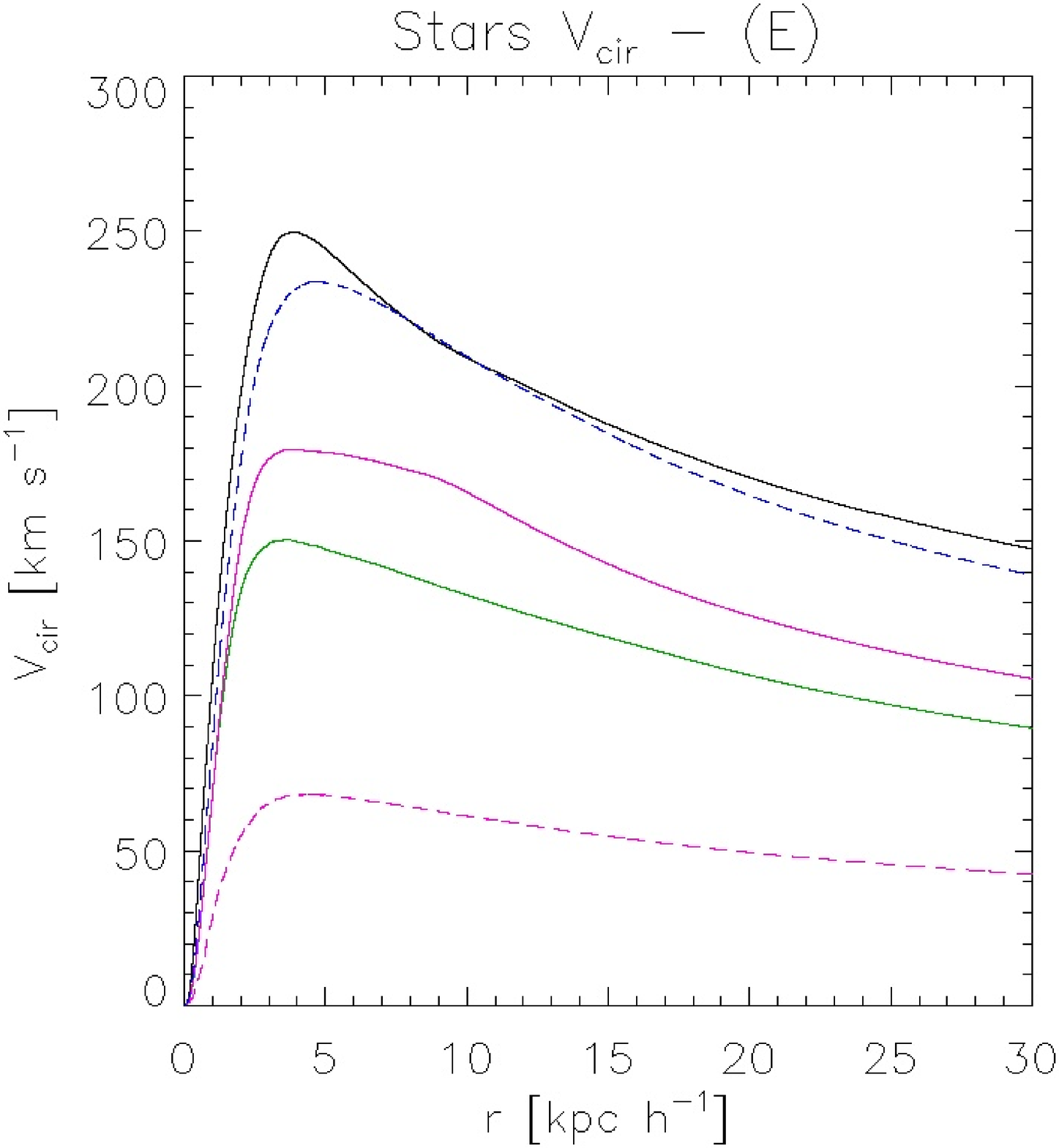}
 \caption{
Dark matter profiles  (left panel) and  stellar  circular velocity (right panel)  for Run NF (black lines), Run E-1 (green continuous lines), Run E-0.7 (pink continuous lines ), Run E-0.3 (blue  dashed lines) and  Run E-3 (pink  dashed lines).}
   \label{fig1}
\end{center}
\end{figure}

As an example, in Fig. \ref{fig1} we show the dark matter  profiles and stellar circular velocity curves for the runs with different assumed energy per SN ($E$)  as well as for the NF case. As larger values for the SN energy are assumed, stronger winds develop regulating the star formation activity so that  a lower fraction of gas is transformed into stars. This behaviour affects the dark matter profiles.
Particularly,  in the extreme unrealistic value of $E=3\times10^{51}$ ergs   where the lowest stellar mass systems is allowed to formed, we obtained the less concentrated  DM profile. We also note that this system is the most dark matter dominated one in the central region as indicated by the total to stellar mass ratio $M_t/M_s$ (Table ~\ref {tab1}). And it shows the lowest stellar circular velocity (Fig. \ref{fig1}, rigth-hand panel).
Interestingly, for Run E-0.7 the dark matter density profile is the most concentrated one. In this case as it can be seen in the right-hand panel of Fig. \ref{fig1} this system shows the flattest stellar mass velocity curve, reflecting the presence of an important disk component.

These results suggest that the amount of stars and their structural distributions are both important factors that
could  be related to the resulting shape of the dark matter profiles.
Similar conclusions can be reached from the analysis of the results for the other two feedback parameters, $F$ and $C$ as it can
be deduced from  Table \ref{tab1}. In the case of the $F$ parameter the response of the dark matter to the variation of $F$ is not linear because the disk structural parameters also change in a more complex way with the strength of the feedback (S08). Also note that compared to the NF case, the feedback runs have lower final stellar masses. For these MW type systems, the fitting of the Einasto profile shows that the dark matter haloes
are in general more concentrated  when feedback is included than in the NF run. In the case of the star formation efficiency , we found that as the $C$  parameter increases, the stellar  mass formed decreases
 and the central systems become more dark matter dominated with more concentrated dark matter  profiles.
The physical reasons of these behaviours will be discussed in detail by Pedrosa et al. (2008, in preparation).

\begin{figure}[h]
\begin{center}
 \includegraphics[width=2.4in]{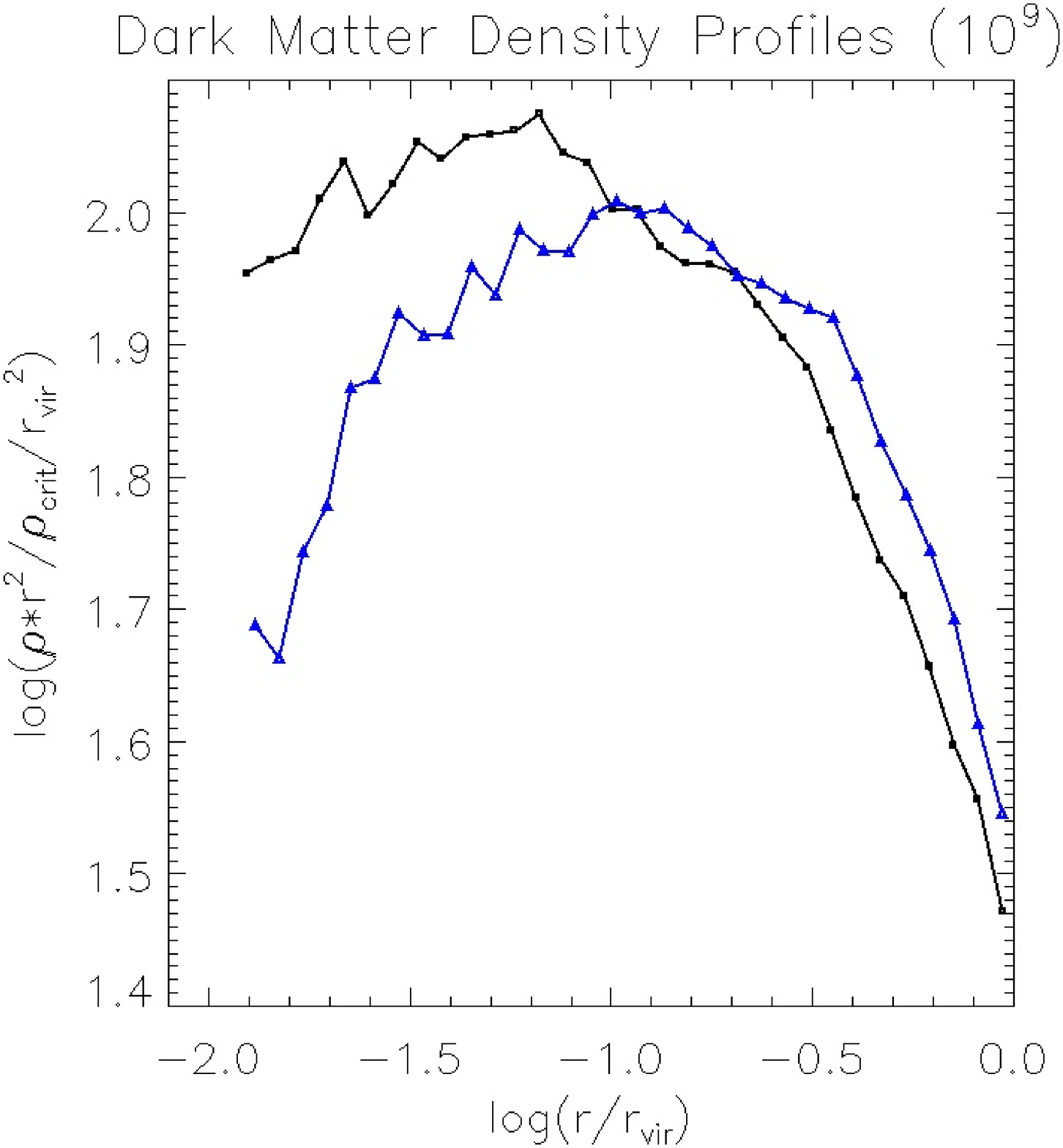}
 \caption{Dark matter density profiles for the scaled-down system of  mass  $10^{9} h^{-1}\ M_{\odot}$. The black line corresponds to the no feedback case while the blue line to a simulations with an energy per SN value of $0.7 \times 10^{51}$ ergs.}
   \label{fig3}
\end{center}
\end{figure}

In order to investigate the dependence of the SN feedback effects on  mass, we studied  a scaled-down system of $10^{9} h^{-1}\ M_{\odot}$,  with the same feedback parameters as in run E-0.7. We have also run the NF version of this small system. For this small galaxy, SN feedback produces a strong decrease in the star formation activity compared with that of the NF case. In this shallow potential well, violent winds  are able to blow out important fraction of  baryons, preventing the  gas from condensating and  forming stars after an early starburst (S08).
 As a consequence, the dark matter  profile shows a strong decrement in the central regions as can be seen in Fig. \ref{fig3}.
However, this result would need to be confirmed with higher resolution simulations.

\begin{figure}[h]
\begin{center}
\includegraphics[width=2.4in]{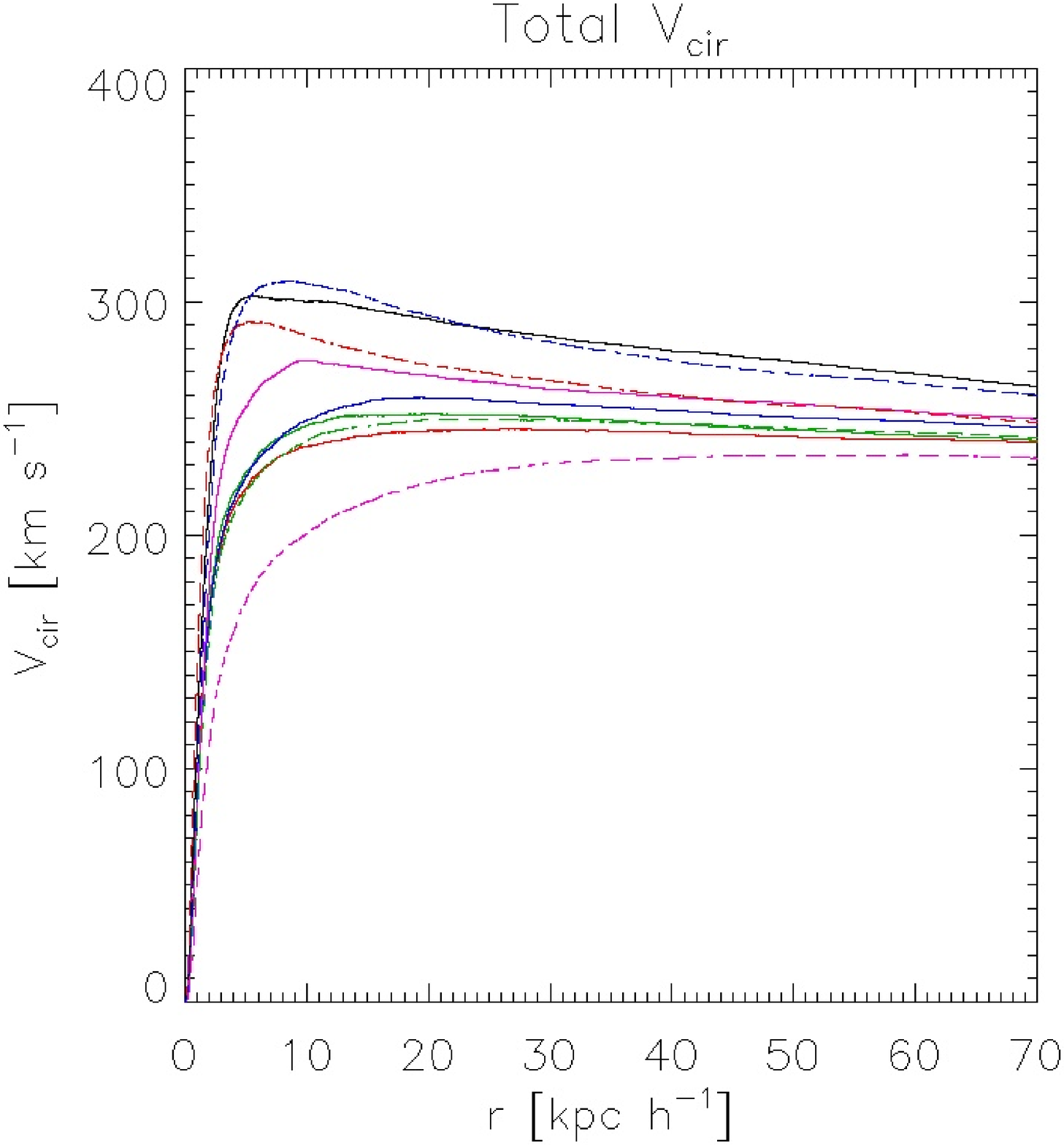}
 \includegraphics[width=2.4in]{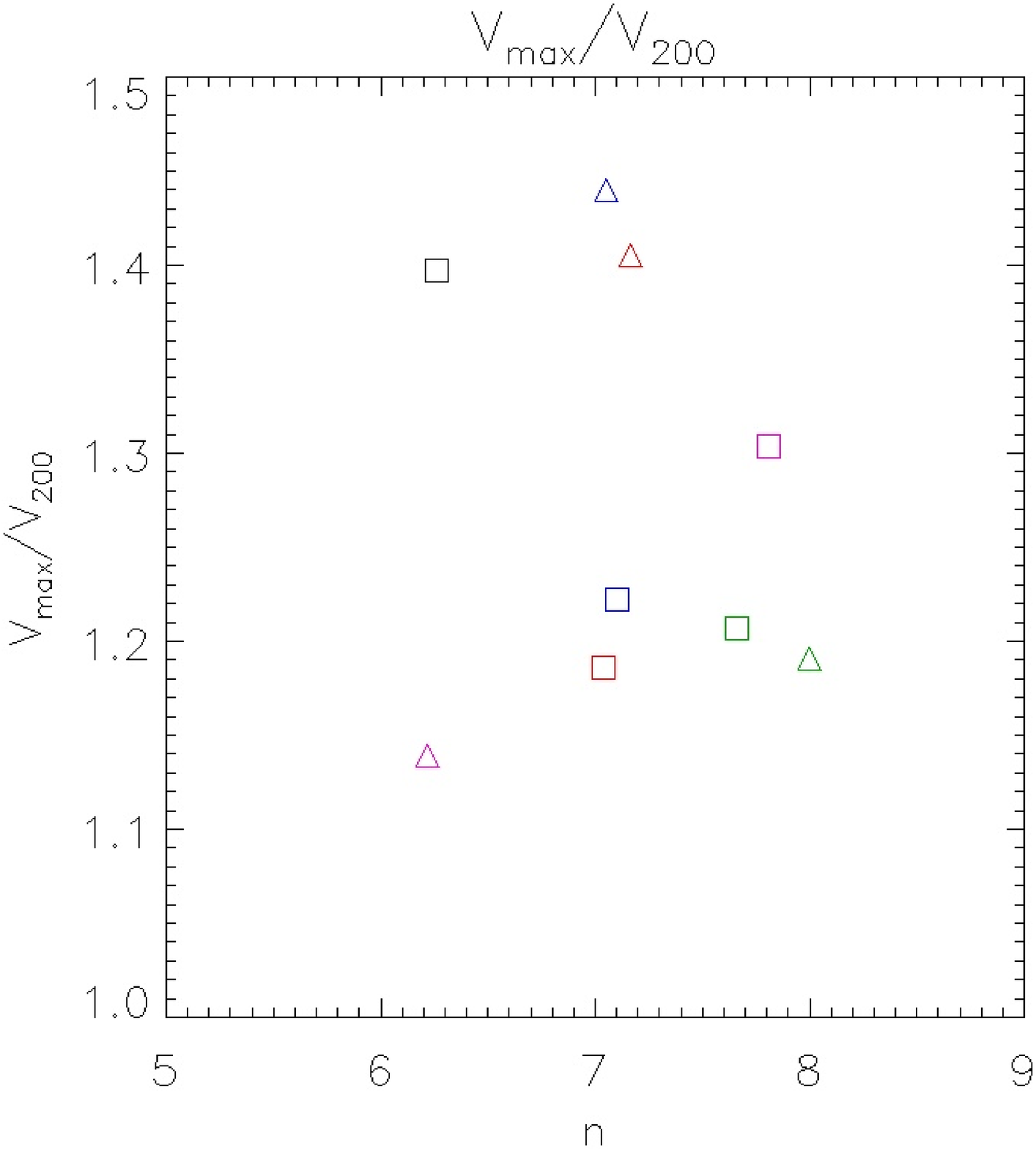}
 \caption{Total circular velocities (left panel) and  cicular velocity maximum to virial velocity ratio as a function  the $n$ parameter (right panel)  for Run NF (black squares and lines), Run F-0.3 (red squares and continuous lines),  Run F-0.5 (green squares and continuous lines), Run F-0.9 (blue squares and continuous lines ), Run E-0.7 (pink squares and continuous lines ), Run C-0.01 (red triangles and dashed lines ) Run C-0.5 (green triangles and dashed lines ), Run E-0.3 (blue  triangles and dashed lines) and  Run E-3 (pink triangles and
dashed lines).}
   \label{fig4}
\end{center}
\end{figure}

 A well-known problem of numerical simulations was the inability to produce flat
rotation curves comparable to that of the Milky Way because of the
catastrophic concentration of baryons at the central region. In
order to evaluate if our simulations are able to overcome this
problem for some combinations of SN and SF parameters, we
calculated the total circular velocity ($V_{\rm c}$).
$V_{\rm c}$
reflects the combined contribution of the dark matter and the baryons inside the central regions.
Spiked $V_{\rm c}$ are found for systems with high central baryonic concentrations (Run NF, Run E-0.3, Run C-0.01), for which the SN feedback was not turned on or was inefficient to regulate the star formation activity.
There are two important effects due to SN feedback as it can be seen in Fig.\ref{fig4}. On one hand, as the  disk component gets better defined, the $V_{\rm c}$   becomes flatter than in the NF case. On the other hand, when systems become more dark matter dominated, as in the extreme case of very strong feedback (Run E-3), the $V_{\rm c}$ grows very smoothly with radius.
To quantify this effect and to compare our results with observations, we estimated the ratio between the maximum of the total circular velocity and the virial velocity.
 In the right panel of Fig. ~\ref{fig4} we show this ratio  as a function of $n$ (shape parameter of the halo). For certain combinations of the SF and SN parameters we obtained ratios where $Vc \approx V_{vir}$, in agreement with observational  results  and theoretical expectations (\cite[Dutton et al. (2008)]{Dutton08} and references therein). We found that this is valid in the cases where the SN feedback was very efficient in blowing out the gas or when an important disk component was able to form.

\section{Conclusions}

This first analysis of the dark matter profiles shows that the dark matter distribution is affected by the way baryons are assembled onto the potential well. As reported by S08, we see that SN feedback prevents the catastrophic gas collapse and foster the formation of extended disk components. In response to this the DM modifies its distribution according to both the final baryonic mass
gathered at the centre and its morphology.
Our findings suggest that the response of the dark matter haloes to feedback  depend on  a combination of galaxy properties such as the amount of stellar mass concentrated at the centre, its shape, and probably the bursty characteristics of the star formation rate. A detailed study is being carried out by Pedrosa et al. (2008 in preparation) in order to gain insight in how these different factors might combined themselves to produce the detected changes in the dark matter profiles and there relative efficiencies.
Also because the strength of the feedback depends on the potential well of the systems, dwarf-type systems could experience more violent mass losses which might produce the flattening of their potential well.
Currently we are studying the history of merger trees of these systems to analyse how the dark matter is affected as the systems are assembled. We are also running higher resolution simulations to check for possible numerical effects.

\end{document}